\documentclass[aps,prl,twocolumn,showpacs,preprintnumbers,amsmath,amssymb]{revtex4}

\usepackage{graphicx}
\usepackage{dcolumn}
\usepackage{bm}
\usepackage{epstopdf}

\begin{document}

\title{Exploring correlated 1D Bose gases from the superfluid to the Mott-insulator state by inelastic light scattering}

\author{D.~Cl\'ement$^{*}$, N.~Fabbri, L.~Fallani, C.~Fort and M.~Inguscio}
\affiliation{LENS, Dipartimento di Fisica,
Universit\`a di Firenze and INFM-CNR, via Nello Carrara 1, I-50019 Sesto Fiorentino (FI), Italy.}

\begin{abstract}
We report the Bragg spectroscopy of interacting one-dimensional Bose gases loaded in optical lattices across the superfluid to Mott-insulator phase transition. Elementary excitations are created with a non-zero momentum and the response of the correlated 1D gases is in the linear regime. The complexity of the strongly correlated quantum phases is directly displayed in the spectra which exhibit novel features. This work paves the way for a precise characterization of the state of correlated atomic phases in optical lattices.
\end{abstract}


\maketitle

Cold atomic gases loaded in optical lattices have been routinely used during the past few years as a versatile and powerful experimental system to study many physical problems \cite{BlochRMP2008}. In particular, they allow to realize and manipulate strongly correlated quantum phases \cite{GreinerNature2002,FallaniPRL2007,MottFermion} and constitute a promising candidate for implementing quantum information processing and quantum simulation schemes \cite{JakschAnnPhys2005}. To achieve those goals, a corner stone consists in a precise characterization of the correlated gaseous phases. 

As in similar problems of condensed-matter, the presence of strong correlations makes it hard to draw a complete picture, both from the experimental and theoretical point of view. The implementation of spectroscopic probes, such as angle-resolved photo-emission spectroscopy for high-Tc superconductors where electron-electron correlations play a major role \cite{DamascelliRMP2003}, are crucial. For the Mott-insulator phase created with trapped atomic gases, experiments have demonstrated the existence of a gap in the spectrum \cite{GreinerNature2002,StoferlePRL2004} and have investigated the shell structure \cite{ShellStructure}, the spatial order via noise correlation techniques \cite{FollingNature2005} and the suppression of compressibility \cite{MottFermion}. Yet an experimental probe giving direct access to important information, such as the temperature or the elementary excitations on which the dynamical properties of the many-body system depend, is still missing. It has been proposed that inelastic light scattering (Bragg spectroscopy) performed at non-zero momentum in the linear response regime could provide such a tool \cite{OostenPRA2005,ReyPRA2005,PupilloPRA2006,HuberPRB2007,MenottiPRB2008}.

In this letter we report the measurement of the linear response of interacting one-dimensional (1D) gases across the superfluid (SF) to Mott-insulator (MI) transition. The 1D gases are loaded in an optical lattice whose amplitude drives the atomic sample from a SF to an inhomogeneous MI state. The elementary excitations are created at a non-zero momentum using two-photon Bragg transition. The presence of an additional mode to the phonon mode in the SF state is suggested. In the inhomogeneous MI state, multiple resonances are observed which give information about the particle-hole excitation energy, the inhomogeneity of the trapped system and which exhibit novel features at low energies that could be related to the temperature of the atomic sample. From the continuous modification of the spectra we also get quantitative information on the critical lattice amplitude for entering the MI phase.

\begin{figure}[ht!]
\begin{center}
\includegraphics[width=\columnwidth]{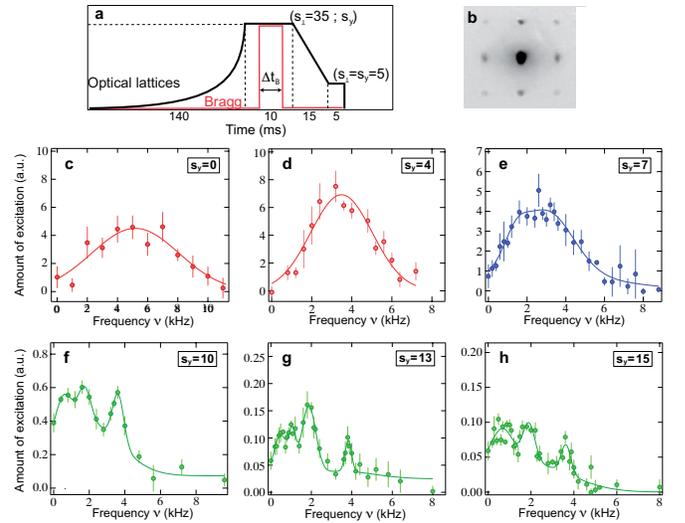}
\end{center}
\caption{\textbf{a} Experimental sequence: after loading the 3D BEC in a 3D optical lattice ($s_{\perp}=35$,$s_{y}$) with an exponential ramp (time constant 30 ms), the 1D gases are excited with the Bragg beams for a time $\Delta t_{\mathrm{B}}$. The optical lattice is then linearly ramped down to $s_{\perp}=s_{y}=5$ in $15~$ms. After a waiting time of $5~$ms, both the optical lattices and the magnetic trap are abruptly switched off and an absorption image is taken after a time-of-flight of $20~$ms. \textbf{b} Typical absorption image in the absence of the Bragg excitation. \textbf{c-h} Spectra in the lowest energy band measured across the SF to MI transition. Lines are guides to the eye. Note the drop in the amplitude of the response (vertical scale).}
\label{Fig1}
\end{figure}

We produce an array of independent 1D gases by adiabatically loading a three-dimensional (3D) Bose-Einstein condensate (BEC) of $^{87}$Rb atoms ($N\simeq 1.5 \times 10^5$, chemical potential $\mu_\mathrm{3D}\simeq 740~$Hz) in two orthogonal optical lattices ($\lambda_{L}=2 \pi/q_{L}=830~$nm) at a large amplitude $V_{\perp}=s_{\perp}E_{\mathrm{R}}=35E_{\mathrm{R}}$ where $E_{\mathrm{R}}=h^2/2 m \lambda_{L}^2$, $h$ being the Planck constant and $m$ the atomic mass. The 1D gases can be considered independent since the resulting tunneling rate in the transverse plane is $0.75~$Hz much smaller than the inverse time scale of the experiment. A crucial quantity indicating the regime in which a 1D gas lies is the ratio of the mean-field interaction energy to the kinetic energy needed to correlate particles on a distance $1/n_{\mathrm{1D}}$,  $n_{\mathrm{1D}}$ being the atomic density \cite{GiamarchiBook}. This ratio is $\gamma= m g_{\mathrm{1D}}/\hbar^2 n_{\mathrm{1D}}$, $g_{\mathrm{1D}}$ being the inter-atomic coupling in the 1D gas \cite{Petrov}. For $\gamma \ll 1$, the 1D gas is in the mean-field regime while for $\gamma \gg 1$ it enters the Tonks-Girardeau regime \cite{GirardeauJMP1960}. In our experiment $\gamma \simeq 0.6$ \cite{MeanGamma} so the correlations in the 1D gases are stronger than in the mean-field regime. To drive the 1D gases from a SF state to a MI state, a longitudinal optical lattice ($\lambda_{L}=830~$nm) with an amplitude $V_{y}=s_{y}E_{\mathrm{R}}$ is added along the axis of the 1D gases. 

The spectroscopic study of the 1D gases is performed using a two-photon Bragg excitation \cite{DavidsonRMP}. Once the 1D gases are loaded into the optical lattices, two off-resonant laser beams ($\lambda_{B}=780~$nm, red-shifted by $250~$GHz from the atomic transition), detuned from each other by a tunable frequency difference $\nu$, are shone onto the atoms during a time $\Delta t_{\mathrm{B}}=3$ to $6~$ms. The momentum transfer $q_{0}$ by the Bragg excitation depends on the angle between the two beams while the energy transferred is given by the frequency difference $\nu$ between them. The angle between the Bragg beams has been set up to transfer a momentum along the $y$ direction smaller but close to the edge of the first Brillouin zone where the response in the Mott state is predicted to be the largest \cite{OostenPRA2005,ReyPRA2005}. The momentum transfer $q_{0}$ along $y$ is kept constant to $q_{0}=0.96(3) q_{L}$ \cite{CalibBraggMom}. The spectra are measured by tuning $\nu$ and monitoring the response of the 1D gases. The experimental sequence is described on Fig.\ref{Fig1}\textbf{a}. From the interference density pattern after expansion (Fig.~\ref{Fig1}\textbf{b}), we fit the central peak with a Gaussian function from which we extract the RMS width $\sigma$. We have checked experimentally that the increase of $\sigma$ induced by the Bragg beams is proportional to the energy transferred \cite{DetailledPaper}. The response of the 1D gases lies in the \emph{linear regime}: on resonance (in the SF and in the MI regime) $\sigma$ increases linearly both with the Rabi frequency $\Omega_{\mathrm{B}}/2 \pi=V_{B}/2 h$ and the duration $\Delta t_{\mathrm{B}}$ of the Bragg pulse, $V_{B}$ being the amplitude of the moving lattice created by the Bragg beams; in addition $V_{B}$ is at the one percent level of the amplitude $V_{y}$ of the longitudinal lattice (for all $s_{y}$ values, $V_{B} < 0.15 E_{\mathrm{R}}$). 

Slightly different parameters $(\Omega_{\mathrm{B}},\Delta t_{\mathrm{B}})$ have been used across the SF-MI transition to obtain a good signal-to-noise ratio. In order to compare all the spectra, the measured increase of $\sigma$ is rescaled with the parameter $\Omega_{\mathrm{B}} \Delta t_{\mathrm{B}}$ and with the total atom number \cite{HWHMatomNumber}. This rescaled quantity is referred to as the amount of excitation and its plot versus the relative detuning $\nu$ as the excitation spectrum. Our excitation scheme is different from previous techniques using lattice modulation \cite{StoferlePRL2004,FallaniPRL2007} for which the momentum transfer is zero and the modulation of the lattice amplitude is large ($\simeq25~\%$).

\textit{SF-MI cross-over} $-$ The array of 1D trapped gases undergoes a cross-over from a SF to an inhomogeneous MI state as the amplitude $s_{y}$ of the longitudinal lattice increases \cite{BatrouniPRL2002}. In the absence of the lattice ($s_{y}=0$), the 1D gases are superfluid at low enough temperature, a property we have checked measuring the frequency ratio of the breathing mode to the dipole mode \cite{MoritzPRL2003}. For large $s_{y}$ ($s_{y}>10$), the 1D gases are deep in the inhomogeneous MI state \cite{StoferlePRL2004}. This quantum phase transition appears on the spectra shown on Fig.~\ref{Fig1}: for low $s_{y}$ the response exhibits a single broad shape (see Fig.~\ref{Fig1}\textbf{c-d}) while, for large $s_{y}$ the spectrum shows a complex structure with multiple resonances (see Fig.~\ref{Fig1}\textbf{g-h}). In addition, the amplitude of the response drastically drops as the atomic system enters the insulating regime and goes deeper in the Mott-insulating phase (note the vertical scale on Fig.~\ref{Fig1}\textbf{f-h}).

\begin{figure}[ht!]
\begin{center}
\includegraphics[width=0.9\columnwidth]{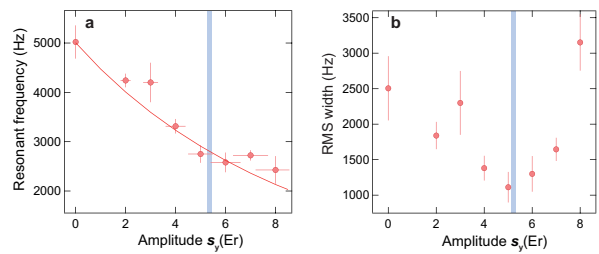}
\end{center}
\caption{\textbf{a} Resonant frequency in the lowest energy band. The solid line corresponds to the mean-field calculation in a single 1D gas for the experimental trapping frequencies and atom number $N_{\mathrm{atom}}=250$. \textbf{b} RMS width of the resonance in the lowest energy band. The blue area is the expected position of the transition for trapped 1D gases extracted from \cite{Rigol2008} and corresponds to $U/J=8.5-9$.}
\label{Fig2}
\end{figure}

The spectroscopy of the 1D gases across the transition allows to mark the boundary between the SF state and the inhomogeneous MI state. In a periodic potential, the energy distribution of the states is captured by the Bloch bands picture. The energy bands of the lattice flatten as the amplitude of the periodic potential is increased. For a momentum transfer close to the edge of the first Brillouin zone $q_{L}$, the flattening of the energy bands implies that the resonant frequency of the transition within the lowest energy band decreases as $s_{y}$ increases. In addition, the tunneling $J$ along the axis of the 1D gases decreases with $s_{y}$ implying a loss of phase coherence and a larger distribution of quasi-momenta populated. Approaching the phase transition, the spectrum width becomes equal to the energy width of the band and decreases with $s_{y}$ as the band flattens. These two effects - the decrease of the resonant frequency and the width of the transition within the lowest energy band - hold for the dephased 1D gases before the phase transition. Once a Mott domain appears, new resonances are expected within the lowest energy band, enlarging the spectrum since the excitation frequencies of the MI state ($\simeq U$ and $\simeq 2U$, see below) are larger than but close to that of the SF state. We have fitted the lineshape of the spectrum corresponding to the transition within the lowest band with a single Gaussian function as the system approaches the phase transition. The resonant frequency and the width are plotted in Fig.~\ref{Fig2}. The width clearly exhibits a minimum which we attribute to the appearance of a MI domain. This position, in the range $U/J=8-10$ (see Eq.~\ref{eqBHH}), is in agreement with recent Monte-Carlo simulations predicting $U/J \sim 8.5-9$ for trapped 1D gases \cite{Rigol2008}. 

\begin{figure}[ht!]
\begin{center}
\includegraphics[width=0.9\columnwidth]{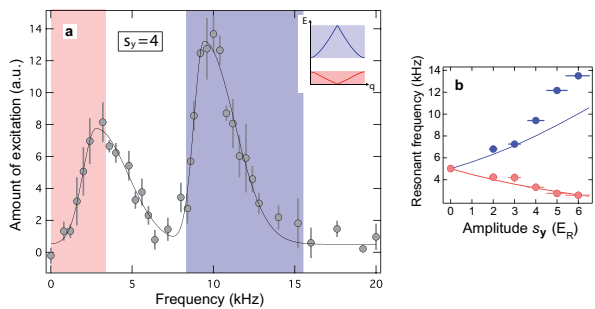}
\end{center}
\caption{\textbf{a} Spectrum of 1D gases in the SF regime ($s_{y}=4$). The solid line is a guide to the eye. The shaded areas correspond to the width of the two lowest bands as shown in the inset. \textbf{b} Resonance frequencies of the first two bands in the SF regime. The lines are mean-field calculations for a 1D BEC with the experimental parameters ($N_{\mathrm{atom}}=250$).}
\label{Fig3}
\end{figure}

\textit{Correlated superfluid regime} $-$ In any periodic system, an excitation at a given quasi-momentum is associated to multiple resonances corresponding to the different energy bands. We restrict our discussion to the excitations towards the first two bands as shown on Fig.~\ref{Fig3}\textbf{a}. In the absence of the lattice ($s_{y}=0$) and for the regime of interactions of the 1D gases ($\gamma \simeq 0.6$), the resonant frequency of the Lieb-Liniger model is indistinguishable from the mean-field solution within our experimental resolution \cite{GiamarchiGroup}. Therefore the modification of the spectrum compared to the mean-field regime mainly consists in the appearance of a tail at low energies \cite{CauxPRA2006}. We compare the Bogoliubov band calculation in the periodic potential with the measured resonant frequencies. We find a good agreement for the lowest energy band while the resonant energy of the transition towards the second band is shifted towards higher energies as the atomic sample approaches the SF-MI transition (see Fig.~\ref{Fig3}\textbf{b})). 

As stated before the presence of correlations in the 1D gases should manifest itself on the shape of the spectra rather than the peak resonant frequency. At $s_{y}=0$ the width we measure ($2.5~$kHz, see Fig.~\ref{Fig1}\textbf{c}) is much larger than the experimental resolution (the spectrum width of the 3D BEC being $0.75(9)~$kHz for the same parameters of the Bragg excitation). We attribute this observation to the presence of correlations and thermal effects peculiar to the beyond mean-field 1D gases \cite{ComingPaper1Dgas}. For $s_{y}>0$, the shape of the spectra in the SF phase becomes asymmetric towards high energies (see Fig.~\ref{Fig3}\textbf{a}). For comparison we have performed the spectroscopy of a 3D BEC loaded in an optical lattice at $s_{y}=6$ measuring a symmetric profile. Numerous theoretical works predict that in addition to the usual phonon mode, an extra gapped mode is expected in strongly correlated superfluids whose amplitude is maximum at the edge of the Brillouin zone \cite{SenguptaPRA2005,HuberPRB2007,MenottiPRB2008}. The momentum transfer $q_{0}=0.96 q_{L}$ is thus appropriate for this observation. The presence of two excitation modes not resolved could support the asymmetry of the spectrum. On the contrary we can exclude other possible explanations. First, the asymmetric shape does not correspond to the algebraic tail of correlated 1D Bose gases which is expected towards low energies \cite{CauxPRA2006}. Second, assuming that the longitudinal lattice dephases the 1D gases implies a large quasi-momentum population that could support the asymmetric shape only for the transition towards the second band (see shaded areas on Fig.~\ref{Fig3}\textbf{a}).

\begin{figure}[ht!]
\begin{center}
\includegraphics[width=0.9\columnwidth]{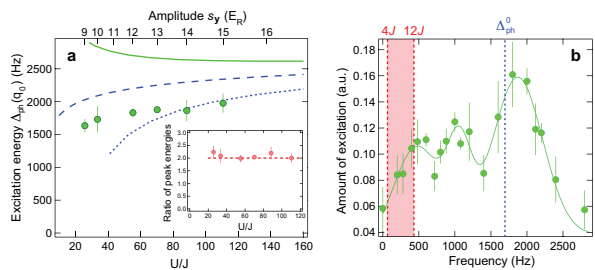}
\end{center}
\caption{\textbf{a} Points correspond to $\Delta_{\mathrm{ph}}(q_{0})$ measured in the inhomogeneous Mott phase. The green solid line is $\Delta_{\mathrm{ph}}(q_{0})$ calculated for a homogeneous MI state at $T=0$ with $n_{0}=3$, the dotted blue line is $\Delta_{\mathrm{ph}}^0$ and the dashed blue line is $U$. Inset: ratio between the two resonant energies in the Mott phase. \textbf{b} Low energy part of the spectrum measured at $s_{y}=13$. The green solid line is a guide to the eye. The shaded area corresponds to $4J-12J$ and the vertical dotted line is $\Delta_{\mathrm{ph}}^0$.}
\label{Fig4}
\end{figure}

\textit{Inhomogeneous Mott regime} $-$ The response of the inhomogeneous MI phase exhibits a complex structure with multiple resonances (see Fig.~\ref{Fig1}\textbf{f-h}) that we now describe. For large amplitudes $s_{y}$ the 1D gases can be described by the Bose-Hubbard Hamiltonian \cite{JakschPRL1998},
\begin{equation}
H=-J \sum_{<i,j>} (a_{i}^{\dag}a_{j}+ \mathrm{h.c.}) + \frac{U}{2} \sum_{i} n_{i} (n_{i}-1) + \sum_{i} \epsilon_{i} n_{i}.\label{eqBHH}
\end{equation}
Here $a_{i}^{\dag}$, $a_{i}$ are the creation and annihilation operator of one boson at site $i$ and $n_{i}=a_{i}^{\dag}a_{i}$ is the particle number operator. The on-site interaction energy is given by $U$ and the next neighbor hopping amplitude by $J$. $\epsilon_{i}$ is the energy offset experienced by an atom on site $i$ due to the presence of the harmonic confining potential. In the homogeneous MI state, the lowest excitation is a particle-hole (ph) excitation which has a minimum energy cost at zero momentum transfer $\Delta_{\mathrm{ph}}^0=U \sqrt{1-4(2n_{0}+1)J/U+(2J/U)^2}$ \cite{HuberPRB2007} with $n_{0}$ being the filling factor ($n_{0}=3$ at the center of our 1D gases). $\Delta_{\mathrm{ph}}^0$, the so-called gap of the MI state, is smaller than $U$ and asymptotically reaches $U$ for large ratio $U/J$. We observe two peaks at frequencies larger than the gap $\Delta_{\mathrm{ph}}^0$. Since the excitations of a SF or a normal gas are expected below this threshold, we attribute these two peaks to the MI domains. The measured frequency of the lowest of these two peaks corresponds to the energy $\Delta_{\mathrm{ph}}(q_{0})$ for exciting a particle-hole at the finite momentum $q_{0}$ and it is plotted on Fig.~\ref{Fig4}\textbf{a} as a function of the ratio $U/J$. The frequency of the second peak is twice the frequency of the first one with their ratio plotted in the inset of Fig.\ref{Fig4} \textbf{a}. Since the response of the atomic gases to the Bragg excitation lies in the linear regime, the peak at the energy $2 \Delta_{\mathrm{ph}}(q_{0})$ does not come from non-linear phenomena. We attribute it to the inhomogeneity of the MI state due to the presence of the trap and a loading in the optical lattices which would not be fully adiabatic \cite{Inhomogeneity}. Information about the inhomogeneity of the Mott state is thus directly accessible using Bragg spectroscopy. We now compare the particle-hole excitation $\Delta_{\mathrm{ph}}(q_{0})$ measured in the experiment to the case of a homogeneous MI at zero temperature $T=0$ (green solid line in Fig.~\ref{Fig4}\textbf{a}) \cite{HuberPRB2007}. Our measurements of $\Delta_{\mathrm{ph}}(q_{0})$ are well below the prediction for the homogeneous case. This deviation can be related both to the inhomogeneity and the finite temperature in affecting the ph excitation energy for which no complete theoretical predictions exist so far. 

Compared to the spectra using lattice modulation \cite{StoferlePRL2004}, novel excitations are observed in the MI state at low energies for all the range $U/J$ tested (see Fig.~\ref{Fig4}\textbf{b}). Since their frequency is smaller than $\Delta_{\mathrm{ph}}^0$ they can not be attributed to an excitation from the ground state of the Mott state. Nevertheless, excitations are expected at these low frequencies in trapped systems at finite temperature. First, the excitation energy of the SF domains with a filling factor smaller than $n_{0}$ is $\simeq 4 n_{0} J$ for a momentum transfer close to the band edge \cite{PupilloPRA2006}. In our experiment we have a maximum filling factor $n_{0}=3$ and we have plotted the energy range $4J-12 J$ as a shaded area on Fig.~\ref{Fig4}\textbf{b}. Second, a non-vanishing temperature $T$ allows the population of several ph excitations which should be observed in the energy transfer induced by Bragg spectroscopy \cite{ReyPRA2005}. The excitations at energies larger than $12 J$ could be explained by the finite temperature of the system. Yet how to extract a temperature from the position and/or width of these low-energy excitations is not clear and we hope that these experimental measurements will trigger a new interest in working out such a possible relation.

In conclusion we have demonstrated the possibility to monitor, in the \emph{linear regime}, the response of correlated quantum gases to an excitation at \emph{non-zero momentum}. We have studied interacting 1D gases from the SF state across to the inhomogeneous MI state. The complexity of such correlated quantum phases appears in the spectra from which important information is directly accessible. In the future, a systematic investigation of the observed novel features combined with new theoretical inputs should allow to adress open questions regarding those strongly correlated gaseous phases.

We are grateful to M. Modugno for inspiring discussions and a critical reading of the manuscript. We thank K.~M.~R.~van der Stam for early work on the experiment. We acknowledge fruitful comments from R. Citro, C. Menotti, A. Minguzzi, N. Trivedi and all the colleagues of the Quantum Degenerate Group at LENS. This work has been supported by UE contract No. RII3-CT-2003-506350, MIUR PRIN 2007, Ente Cassa di Risparmio di Firenze, DQS EuroQUAM Project, NAMEQUAM project and Integrated Project SCALA. 

\bibliography{apssamp}

\begin{thebibliography}{99}
\bibitem{CalibBraggMom} The momentum transfer $q_{0}$ is calibrated using a 3D BEC by monitoring the distance between the diffracted and non-diffracted clouds after a time-of-flight and by measuring the resonance frequency of the Bragg excitation.
\bibitem{DetailledPaper} These properties will be detailled in a future publication.
\bibitem{HWHMatomNumber} In the absence of Bragg excitation, the dependence of $\sigma$ with the total atom number has been calibrated from which the value $\sigma$ for each image is normalized.
\bibitem{ComingPaper1Dgas} The spectra of the 1D gases as a function of $s_{\perp}$ have been studied and will be the subject of a future publication.
\end{thebibliography}

\end{document}